\documentclass[11 pt]{iopart}

\usepackage{graphicx,color}
\usepackage{dcolumn}
\usepackage{epsfig}
\usepackage{iopams}
\usepackage{cite}

\usepackage{amsfonts}
\usepackage{euscript}
\usepackage{amssymb}
\usepackage{amsthm}
\usepackage{newlfont}
\usepackage{mathrsfs}
\usepackage{subfigure}

\newcommand{\ie}{\textit{i.e.}}

\newcommand{\id}{\textrm{d}}

\def\bea{\begin{eqnarray}}
\def\eea{\end{eqnarray}}
\def\ba{\begin{array}}
\def\ea{\end{array}}

\def\la{\langle}
\def\ra{\rangle}

\begin{document}

\title[Exact stationary state of a three-state run-and-tumble particle]{Exact stationary state of a run-and-tumble particle with three internal states in a harmonic trap}
\author{Urna Basu$^1$, Satya N. Majumdar$^2$, Alberto Rosso$^2$, Sanjib Sabhapandit$^1$, Gr\'{e}gory Schehr$^2$}

\address{$^1$Raman Research Institute, Bengaluru 560080, India}
\address{$^2$LPTMS, CNRS, Univ. Paris-Sud, Universit{\'e} Paris-Saclay, 91405 Orsay, France}

\begin{abstract}
We study the motion of a one-dimensional run-and-tumble particle with three discrete internal states in the presence of a harmonic trap of stiffness $\mu.$ The three internal states, corresponding to positive, negative and zero velocities respectively, evolve following a jump process with rate $\gamma$. We compute the stationary position distribution exactly for arbitrary values of $\mu$ and $\gamma$ which turns out to have a finite support on the real line. We show that the distribution undergoes a shape-transition as $\beta=\gamma/\mu$ is changed.  For $\beta<1,$ the distribution has a double-concave shape and shows algebraic divergences with an exponent $(\beta-1)$ both at the origin and at the boundaries. For $\beta>1,$ the position distribution becomes convex, vanishing at the boundaries and with a single, finite, peak at the origin. We also show that for the special case $\beta=1,$ the distribution shows a logarithmic divergence near the origin while saturating to a constant value at the boundaries. 
\end{abstract}

\section{Introduction}

Recent years have seen a surge of interest in the study of active matter and active particles. 
The term `active particle' refers to a class of self-propelled particles which can generate dissipative directed motion by consuming energy directly from their environment \cite{Romanczuk,soft,BechingerRev,Ramaswamy2017,Marchetti2017,Schweitzer}. Examples of active matter can be found in nature at all length scales, ranging from micro-organisms like bacteria \cite{Berg2004,Cates2012} to granular matter \cite{gran1,gran2}, flock of birds \cite{flocking1, flocking2} and fish-schools \cite{Vicsek,fish}. Apart from a diverse set of novel collective behaviours like clustering \cite{cluster1,cluster2,evans}, motility induced phase separation \cite{separation1, separation2, separation3}, and absence of well defined pressure \cite{Kardar2015}, active particles show many intriguing features even at the single particle level. One such interesting feature is that, in the presence of external potentials and confining boundaries, active particles show very different behaviour than their passive counterparts, including non-Boltzmann stationary state, clustering near the boundaries of the confining region \cite{Solon2015,Potosky2012,ABP2019,RTP_trap,Malakar2019} and unusual relaxation and persistence properties \cite{RTP_free,ABP2018, Singh2019}.  There have been numerous recent studies focusing on the behaviour of active particles in the presence of  external potentials and confinements, both theoretical \cite{Franosch2016,Das2018,Caprini2019,Sevilla2019} and experimental \cite{Hagen2014,Takatori,Deblais2018,Dauchot2019}.

The theoretical attempts to characterise the behaviour of active particles focus on studying simple models of such systems. Run-and-tumble particle (RTP) is one of the most studied models of an active particle. An RTP is an overdamped particle which moves with a constant speed $v_0$, or `runs,' along the direction of an internal `spin' degree of freedom. The orientation of the spin can change randomly resulting in a sudden change, or `tumble,' in the direction of motion of the particle. The simplest example is an RTP moving in one spatial dimension with two possible values of the spin $\sigma = \pm 1.$ In this case, the particle moves with velocity $v_0$ or $-v_0;$ the reversal of direction occurs stochastically with rate $\gamma,$ with the flipping of the spin $\sigma \to - \sigma.$  In the presence of an external potential $U(x),$ the position $x(t)$ of this two-state RTP evolves according to the  Langevin equation,
\bea
\dot x = f(x) + v_0 \sigma(t) \label{eq:RTP_2st}
\eea
where $f(x)= - U'(x)$ is the deterministic force acting on the particle. The spin variable $\sigma$ plays the role of the noise, its dichotomous nature giving rise to the `activity'. In fact, it is clear from the auto-correlation $\la \sigma(t) \sigma(t') \ra = e^{-2 \gamma |t-t'|}$  that $\sigma(t)$ is a coloured noise with a finite memory, characterised by the persistence time $\tau = (2\gamma)^{-1}.$  Despite the apparent simplicity of the model, the two-state RTP shows a lot of intriguing features typical to active particles including non-Boltzmann stationary distribution\cite{RTP_trap, RTP_free}. 

For any confining potential, the stationary position distribution of a two-state RTP is known exactly, and is given by, 
\bea
P_\textrm{st}(x) \propto \frac {1}{v_0^2-f^2(x)}\exp{\bigg[2 \gamma \int_0^x \id y \frac{f(y)}{v_0^2-f^2(y)}\bigg]} \label{eq:Pst_2st}
\eea
up to a normalization constant. %This distribution has a finite support on the real line, which depends on the particular potential. 
The above result was first obtained long ago in the context of quantum optics \cite{q-optics1,q-optics2,q-optics3,q-optics4}, and later to study the role of coloured noise in dynamical systems \cite{colored}. More recently, it has been re-derived in the context of active particles \cite{Kardar2015, RTP_trap}. In particular, the stationary distribution \eref{eq:Pst_2st} has been analysed for specific confining potentials of the type $U(x) \propto |x|^p$ with $p>0$ in Ref. \cite{RTP_trap}. The case $p=2$ corresponds to a harmonic potential which is of particular interest, not only from theoretical but also from an experimental point of view \cite{Takatori,Dauchot2019}. For a harmonic potential $U(x) = \mu x^2/2,$ the stationary distribution \eref{eq:Pst_2st} simplifies to,
\bea
P_\textrm{st}(x)= \frac{2 \mu}{4^{\beta}B(\beta, \beta)v_0} \left [1- \left(\frac{\mu x}{v_0} \right)^2 \right]^{\beta -1} \label{eq:2st_harmonic}
\eea
where $\beta = \gamma/\mu$ and $B(u,v)$ is the beta-function. This distribution is symmetric in $x$ and has a finite support in the region $-\frac{v_0}{\mu} \le x  \le \frac{v_0}{\mu}.$ Consequently,  the particle is confined within this region in the stationary state.  This stationary position distribution shows an interesting shape-transition as a function of $\beta.$ For $\beta > 1$ the distribution is convex shaped, with a peak at the origin $x=0$ and $P_\textrm{st}(x)$ vanishing at the boundaries $x = \pm \frac{v_0}{\mu}.$ On the other hand, for $\beta < 1$ $P_\textrm{st}(x)$ has a concave shape with divergences at the boundaries and a minimum at the origin. For $\beta=1,$ the distribution is uniform. Thus by varying $\beta,$ one can observe a transition from a double-peaked (at the boundaries) to a single-peaked distribution. The double-peaked nature of the distribution for $\gamma < \mu$ signifies an `active phase', where the persistence time of the spin-orientation is larger than $\mu^{-1},$ the relaxation time-scale of the potential. On the other hand, $\gamma > \mu,$ \ie, when the persistence time is smaller compared to $\mu^{-1},$ corresponds to a passive phase, where the stationary distribution resembles that of a passive particle in a trap, with a single peak at the centre of the trap. Indeed, in the diffusive limit when $v_0 \to \infty$, $\gamma \to \infty$ but keeping the ratio $v_0^2/2\gamma = D$ fixed, the dynamics of the RTP in the harmonic trap converges to the Ornstein-Uhlenbeck process. This is also exhibited in the stationary state where the distribution in Eq. (\ref{eq:2st_harmonic}) converges to a Boltzmann distribution, which in this case is a simple Gaussian $P_{\textrm{st}}(x) \propto e^{- \frac{\mu}{D}x^2}$.

It is then natural to ask how the stationary distribution changes if the RTP has more than two internal states. In fact, an RTP with many internal degrees have been studied where the internal degrees can take a set of discrete values and evolve following some discrete jump processes \cite{Seifert2016,Maes2018}. 
%On the other hand, RTPs in higher dimensions have also been studied, where the internal spin can assume any orientation.
However, most of these studies are numerical and to the best of our knowledge no analytical results are available for the stationary state of a multi-state RTP in the presence of an external potential.

In this article, we study a run-and-tumble active particle in one spatial dimension with three discrete internal states, with positive, negative and zero velocities, respectively. We show that such a multi-state dynamics naturally arises when one considers an RTP in higher spatial dimensions and project it to one-dimension. We calculate exactly the stationary position probability distribution in the presence of a harmonic potential of strength $\mu$ for arbitrary flip-rate $\gamma$ among the internal states. It turns out that the presence of the zero-velocity internal state leads to a rich behaviour of the position distribution $P(x)$. As in the two-state case, it turns out that the shape of the stationary state distribution is governed by one single parameter
\begin{eqnarray}\label{def_beta}
\beta=\frac{\gamma}{\mu} \;.
\end{eqnarray}
We show that $P(x)$ has a finite support on the real line and undergoes a transition in shape as $\beta=\gamma/\mu$ is varied : For $\beta <1,$  $P(x)$ diverges both at the origin and the boundaries with the same exponent $\beta -1.$ Thus, in this case, the position distribution has a double-concave shape, with three peaks, namely at the boundaries and the origin. For $\beta=1,$ $P(x)$ shows a logarithmic divergence near the origin. On the other hand, for $\beta >1,$ the distribution converges to a finite value at the origin while it vanishes at the boundaries, implying a convex shape with a single peak at the origin (see Fig. \ref{fig:px}).

\section{Model}

\begin{figure}[t]
 \centering
 \includegraphics[width=8 cm]{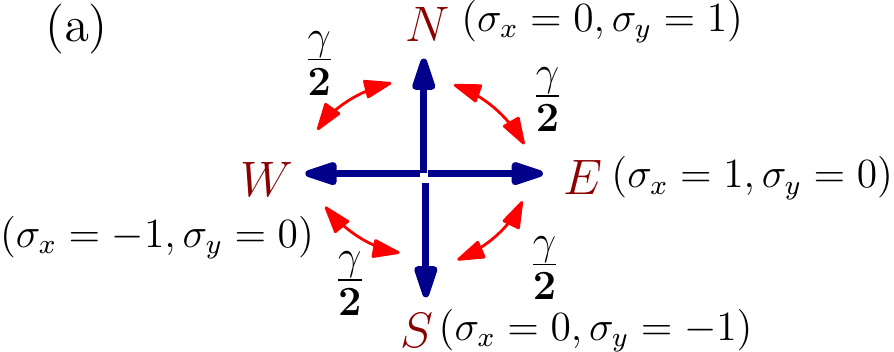} \hspace{1.5 cm}\includegraphics[width=3.5  cm]{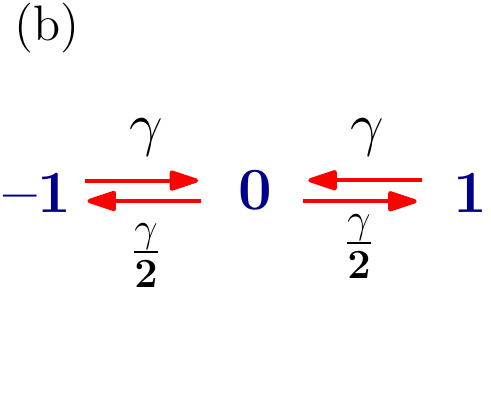} 
 % spin_4st2.pdf: 0x0 pixel, 300dpi, 0.00x0.00 cm, bb=
 \caption{(a) Schematic representation of the jump process through which the orientation $\sigma$ evolves. (b) The equivalent 3-state jump process for $\sigma_x$.}
 \label{fig:4st_3st}
\end{figure}

Our model of a three-state RTP in one-dimension is motivated by a natural ``clock-like'' model for a two-dimensional RTP. 
Let us indeed consider an overdamped particle moving on a two dimensional $(xy)$ plane with an internal orientational degree of freedom or `spin' $\sigma$ associated with it. In the absence of any external potential the particle moves with a constant speed $v_0$ along the direction of $ \sigma,$ which is a unit vector with  four possible discrete orientations, denoted by $E,W,N,S$ (along $\pm x$ and $\pm y$ axes respectively). The spin $\sigma$ evolves in time following a Markov jump process -- its orientation can change via a rotation of $\frac\pi 2$ either clockwise or anti-clockwise, both with rate $\frac \gamma 2.$ This jump process is schematically represented in Fig. \ref{fig:4st_3st}(a). Additionally, we consider an external harmonic potential $U(x,y) = \frac \mu 2(x^2+y^2)$ which exerts a force $f(x,y) = -\nabla U(x,y)$ on the RTP. 

%At any time $t,$ the configuration of the RTP is completely specified by its position $(x(t),y(t))$ and the orientation vector $\sigma(t).$ 
The time-evolution of the position $(x(t),y(t))$ of the RTP can be conveniently expressed in terms of the Langevin equations,
\numparts
\bea
\dot x(t) &=& -\mu x(t) + v_0 \sigma_x(t) \label{eq:2dmodel_x} \\
\dot y(t) &=& - \mu y(t) + v_0 \sigma_y(t) \label{eq:2dmodel_y}
\eea
\endnumparts
where $\sigma_{x,y}(t)$ are components of the spin vector $\sigma(t)$ at any time $t,$ along the $x$ and $y$ axes respectively (see Fig. \ref{fig:4st_3st}(a)). 
%Note that $\sigma_x =1$ for $\sigma=E,$ $\sigma_x=-1$ for $\sigma=W$ and $\sigma_x=0$ for $\sigma=N,S.$ Similarly, $\sigma_y=\pm 1$ and $0$ for $\sigma=N,S$ and $E,W$ respectively (see Fig. \ref{fig:schem}).  

The position probability distribution $\cal P(x,y,t)$ is given by the sum $\cal P(x,y,t) = \sum_{\sigma} \cal P_\sigma(x,y,t)$ where $\cal P_\sigma(x,y,t)$ denotes the probability that the particle has the position $(x,y)$ and orientation $\sigma = E, N, W, S$ at time $t.$ These probabilities evolve according to the Fokker-Planck (FP) equations,
\numparts
\bea
\fl \qquad \frac{\partial}{\partial t} \cal P_E(x,y,t) = \frac{\partial }{\partial x} \bigg[(\mu x - v_0)\cal P_E \bigg] + \frac{\partial }{\partial y} \bigg[\mu y \cal P_E \bigg] + \frac \gamma 2 (\cal P_N+\cal P_S) - \gamma \cal P_E \label{eq:FP_2d1}\\
\fl \qquad \frac{\partial}{\partial t} \cal P_N(x,y,t) =  \frac{\partial }{\partial x} \bigg[ \mu x \cal P_N \bigg] + \frac{\partial }{\partial y} \bigg[(\mu y - v_0) \cal P_N \bigg] + \frac \gamma 2 (\cal P_E+\cal P_W) - \gamma \cal P_N \\
\fl \qquad \frac{\partial}{\partial t} \cal P_W(x,y,t) =  \frac{\partial }{\partial x} \bigg[ (\mu x + v_0)\cal  P_W \bigg] + \frac{\partial }{\partial y} \bigg[\mu y \cal  P_W \bigg] + \frac \gamma 2 (\cal P_N+\cal P_S) - \gamma \cal P_W \\
\fl \qquad \frac{\partial}{\partial t} \cal P_S(x,y,t) =  \frac{\partial }{\partial x} \bigg[ \mu x\cal  P_S \bigg] + \frac{\partial }{\partial y} \bigg[(\mu y + v_0)  \cal P_S \bigg] + \frac \gamma 2 (\cal P_E+\cal P_W) - \gamma \cal P_S \label{eq:FP_2d}
\eea
\endnumparts
where we have suppressed the argument of $P_\sigma$ on the right hand side for the sake of brevity. 
It is hard to find an analytical form of $\cal P(x,y,t)$ as these equations are difficult to solve, even in the stationary state. %Later in Sec. \ref{sec:2d} we will study the stationary position distrubtion $\cal P(x,y)$ using numerical simulations. 

However, it is also interesting to look at the  $x$-process only, governed by Eq.~\eref{eq:2dmodel_x}. This describes an effective one-dimensional RTP where the internal spin $\sigma_x$ has three possible discrete values, $1,0,-1.$ 
As illustrated in Fig. \ref{fig:4st_3st}(a), both $\sigma=N$ and $\sigma=S$ correspond to $\sigma_x=0$ while $\sigma=E$ and $\sigma=W$ corresponds to $\sigma_x=1$ and $\sigma_x=-1,$ respectively. The jump from $\sigma_x=1$ to $0$ can, thus, occur through two different channels ($E \to N$ and $E \to S$), resulting in a jump rate $\gamma$ for $\sigma_x=1 \to \sigma_x=0.$ Similarly, $\sigma_x=-1 \to \sigma_x =0$ occurs with rate $\gamma,$ while $0 \to \pm 1$ occurs with rate $\frac \gamma2$ (since there is only one way to make this transition). This effective 3-state jump process in one-dimension is schematically shown in Fig.~\ref{fig:4st_3st}(b). Let $P_{i}(x,t)$ denote the probability that the RTP is at a position $x$ at time $t$ with $\sigma_x=i.$ The corresponding FP equations read,
%We can now write the FP equation for this effective 1-dimensional process,
\numparts
 \bea
 \frac{\partial}{\partial t} P_{1}(x,t) &=& \frac{\partial }{\partial x} [(\mu x - v_0 ) P_{1}] + \frac\gamma 2 P_0 - \gamma P_1 \label{eq:FP_3st1}\\
 \frac{\partial}{\partial t} P_{-1}(x,t) &=& \frac{\partial }{\partial x} [(\mu x + v_0 ) P_{-1}] + \frac\gamma 2 P_0 - \gamma P_{-1} \label{eq:FP_3stm1}\\
 \frac{\partial}{\partial t} P_{0}(x,t) &=& \frac{\partial }{\partial x} [\mu x P_{0}] + \gamma  (P_1 + P_{-1}) - \gamma P_0. \label{eq:FP_3st0}
 \eea 
 \endnumparts
We note that this set of FP equations can also be obtained from Eqs.~\eref{eq:FP_2d1}-\eref{eq:FP_2d} by integrating both sides over $y$ and then identifying  $P_1(x,t) = \int \id y ~\cal P_E(x,y,t),$ $P_{-1}(x,t) = \int \id y ~\cal P_W(x,y,t),$ and $P_0(x,t)= \int \id y ~[\cal P_N(x,y,t) + \cal P_S(x,y,t)].$

In the presence of the confining harmonic potential, in the long time limit the RTP is expected to reach a stationary state where the left hand side (l. h. s.) of the Eqs.~\eref{eq:FP_3st1} - \eref{eq:FP_3st0} would vanish. The corresponding stationary distributions $P_i(x) = \lim_{t \to \infty} P_i(x,t)$ then satisfy a set of coupled linear differential equations (obtained by putting $\frac{\partial P_i}{\partial t}=0$),
\numparts
\bea
\frac{\id }{\id x} [(\mu x - v_0 ) P_{1}] + \frac\gamma 2 P_0 - \gamma P_1 &=& 0 \label{eq:st_P1} \\
\frac{\id }{\id x} [(\mu x + v_0 ) P_{-1}] + \frac\gamma 2 P_0 - \gamma P_{-1}&=& 0 \label{eq:st_Pm1}\\
 \frac{\id}{\id x} [\mu x P_{0}] + \gamma  (P_1 + P_{-1}) - \gamma P_0&=&0. \label{eq:st_P0} 
\eea
\endnumparts
Our objective is to solve this set of equations to find $P_i(x)$ in the stationary state. \\

\noindent {\bf Boundary Conditions:} To proceed with the solution we first need to specify the boundary conditions for $P_i(x).$ To determine these boundary conditions, we first note that, in the stationary state,  the RTP is confined within a finite region bounded by $x_\pm = \pm v_0/\mu.$ This can be understood easily from the following argument: from the Langevin equation \eref{eq:2dmodel_x} it is clear that %the instantaneous velocity $\dot x$ of the particle is always positive (or, negative) if the particle is in the region $x<x_-$ (or, $x > x_+$), \ie, 
if the particle is outside the region  $[x_-,x_+],$ it always feels a drift towards the origin, irrespective of the value of $\sigma_x.$ As a result, if the particle starts from some initial position $x_0 > x_+,$ or $x_0 < x_-,$ it will eventually reach the region  $[x_-,x_+].$  Consequently, the stationary distribution has a finite support in the region $[x_-,x_+]$ 
and it is zero outside. To solve Eqs.~\eref{eq:st_P1} - \eref{eq:st_P0} then, we need to specify the  boundary conditions at these two points. Let us first look at the behaviour of $P_1(x)$ near $x=x_-.$ During an infinitesimal time increment $\Delta t,$ $P_1(x_-,t)$ evolves as,
\bea
P_1(x_-,t+ \Delta t) = (1- \gamma \Delta t) P_1(x_- -\Delta x,t) + \frac \gamma 2 \Delta t P_0(x_-,t) \label{eq:P1_dt}
\eea
where the first term on the right hand side (r.h.s.) represents the transition when the position of the particle changes by an amount $\Delta x$ during interval $\Delta t,$ and the second term corresponds to the case when $\sigma_x$ changes from $0$ to $1;$ the pre-factors $(1- \gamma \Delta t)$ and $\gamma \Delta t/2$ denotes the probabilities for these two occurrences, respectively. Now, in the stationary state, the probabilities $P_i(x)$ are independent of time, hence, we have from \eref{eq:P1_dt},
\bea \label{Eq:P1}
P_1(x_-) =  (1- \gamma \Delta t) P_1(x_- -\Delta x) + \frac \gamma 2 \Delta t P_0(x_-)
\eea
Moreover, from Eq.~\eref{eq:2dmodel_x} we have, for $\sigma_x=1$ and near $x_-,$ $\Delta x \simeq (- \mu x_- + v_0) \Delta t = 2 v_0 \Delta t >0,$ thus $P_1(x_- - \Delta x) = P_1(x_--2 v_0 \Delta t)$ which vanishes in the stationary state, as the argument $x_- - 2 v_0 \Delta t$ is outside the region $[x_-,x_+]$. Then, taking $\Delta t \to 0$ limit in Eq. (\ref{Eq:P1}), we get $P_1(x_-)=0.$ Using similar arguments for $P_{-1}$ and $P_0$, one finds the full set of boundary conditions to be satisfied by the set of equations \eref{eq:st_P1} - \eref{eq:st_P0},
\bea
P_1(x_-)=0, \; P_{-1}(x_+) =0, \; P_0(x_-) =0, \; P_0(x_+) =0. \label{eq:bc2}
\eea
Note that the behaviour of $P_1(x_+)$ and $P_{-1}(x_-)$ remain unspecified. The set of boundary conditions for $P_1$ and $P_{-1}$ is very similar to the case of 2-state RTP \cite{RTP_trap}. However, as we will see below, the presence of the third state $\sigma_x=0$ leads to a richer behaviour in the present case. \\

\section{Exact Solution} 

The straightforward strategy to solve a set of coupled first order equations like Eqs.~\eref{eq:st_P1} - \eref{eq:st_P0} is to decouple them and find separate equations for $P_i(x).$ However, our primary goal is to find the marginal position distribution of the particle, \ie, the probability that the effective one-dimensional RTP has a position $x,$ irrespective of the spin-orientation $\sigma_x.$ This is given by
\bea
P(x) &=& P_0(x) + P_{1}(x) + P_{-1}(x). \label{eq:P-def}
\eea
In the following we attempt to derive an equation for $P(x)$ using Eqs.~\eref{eq:st_P1} - \eref{eq:st_P0}. To this end, we first define,
\bea
Q(x) = P_{1}(x) + P_{-1}(x),\quad \textrm{and} \quad R(x) = P_{1}(x) - P_{-1}(x).  \label{eq:QR-def}
\eea
It is straightforward to see that in terms of these functions $P$ and $Q$, the four boundary conditions given by Eq.~\eref{eq:bc2} translate to,
\bea
P(x_+) = Q(x_+), \quad \textrm{and} \; P(x_-)=Q(x_-) \label{eq:bc_PQ}
\eea
Note that the boundary conditions of $R(x)$ remain unspecified. 
We proceed by expressing Eqs.~\eref{eq:st_P1} - \eref{eq:st_P0} in terms of these  functions $P$ and $Q.$ For this purpose, we first add equations \eref{eq:st_P1}, \eref{eq:st_Pm1} and \eref{eq:st_P0} to get, 
\bea
\frac{\id}{\id x} \bigg[\mu x P(x) - v_0 R(x)\bigg] =0 \; \Rightarrow \mu x P(x) - v_0 R(x) = C \label{eq:C}
\eea
where $C$ is a constant independent of $x.$ To determine $C,$ we substitute $x=x_+$ in the above equation. Using the definitions of $P$ and $R,$ along with the boundary condition \eref{eq:bc2}, we get, $C = (\mu x_+ - v_0) P_1(x_+) = 0.$ Hence, from Eq.~\eref{eq:C} we have,
\bea
R(x) = \frac{\mu x}{v_0} P(x) \label{eq:PR}
\eea
for all values of $x.$  Now, adding Eqs.~\eref{eq:st_P1} and \eref{eq:st_Pm1} and using Eq.~\eref{eq:PR}, we get, 
\bea
\mu x P'(x) + (\mu - \gamma) P(x) = \mu x Q'(x) + (\mu - 2 \gamma) Q(x) \label{eq:PQ_add}
\eea
where $'$ denotes the derivative with respect to (w.r.t.) the argument of the functions. Next, we subtract Eq.~\eref{eq:st_Pm1} from Eq.~\eref{eq:st_P1} to get, 
\bea
(\mu x)^2 P'(x) + \mu (2 \mu - \gamma) x P(x) = v_0^2 Q'(x). \label{eq:PQ_subtract}
\eea  
Eqs.~\eref{eq:PQ_add} and \eref{eq:PQ_subtract} are two coupled linear differential equations involving $P(x)$ and $Q(x).$ In the following, we use them to get two separate differential equations for $P(x)$ and $Q(x).$ But, first, it is convenient to use a change of variable $z= (\frac{\mu x}{v_0})^2$ with $0 \le z \le 1.$ Let us denote $\tilde P(z)=P(x=v_0\sqrt{z}/\mu)$ and $\tilde Q(z)= Q(x=v_0\sqrt{z}/\mu).$  Eqs.~\eref{eq:PQ_add} and \eref{eq:PQ_subtract} then become,
\bea
2z \tilde P'(z) +(1-\beta) \tilde P(z) &= & 2z \tilde Q'(z) +(1- 2 \beta)\tilde Q(z)\label{eq:PQz1} \\
z  \tilde P'(z) + \left(1- \frac \beta 2 \right) \tilde P(z) &=& \tilde Q'(z) \label{eq:PQz2}
\eea
where $\beta= \gamma /\mu.$ The two boundary conditions in Eq.~\eref{eq:bc_PQ} reduce to a single condition for $\tilde P$ and $\tilde Q,$
\bea
\tilde P(z=1) = \tilde Q(z=1) \;. \label{eq:bc_z}
\eea
As we will see below, this boundary condition is enough to solve the differential equations uniquely. 

%Clearly, the position distribution depends only on $\beta,$ not on $\mu$ and $\gamma$ invidually. 

To get an equation involving $\tilde P(z)$ only, we take derivative of Eq.~\eref{eq:PQz1} w.r.t. $z.$ Then, using Eq.~\eref{eq:PQz2}, we immediately arrive at a second order differential equation,
\bea
\fl \qquad z(1-z)  \tilde P''(z) + \left[\frac{3-\beta}2 - \frac 12(7-3\beta)z \right]\tilde  P'(z) -\left(1- \frac \beta 2 \right)\left(\frac 32 - \beta\right)\tilde P(z)=0 \label{eq:Pz}
\eea
It is straightforward to check that the above equation is in the form of a hypergeometric differential equation,
\bea
z(1-z) \tilde P''(z) +[c_1-(a_1+b_1+1)z] \tilde P'(z)  - a_1b_1\tilde P(z) =0 \label{eq:Hyp}
\eea
with the parameters,
\bea
a_1= 1-\frac \beta2; \quad b_1 = \frac 32-\beta; \quad c_1 = \frac {3 - \beta}2.
\eea
One can also get a similar second order equation for $\tilde Q(z).$
To this end, we first express $P'(z)$ in terms of $\tilde Q(z)$ and $\tilde Q'(z)$, \ie, in a form similar to Eq.~\eref{eq:PQz2}. Multiplying Eq.~\eref{eq:PQz1} by $(1-\frac \beta 2)$ and Eq.~\eref{eq:PQz2} by $(1-\beta),$ and subtracting the latter resulting equation from the former, we get,
\bea
z \tilde P'(z) = (1- 2 \beta)\left(1- \frac \beta 2 \right)\tilde Q - [1- \beta -(2- \beta)z]  \tilde Q'(z) \label{eq:dPz}
\eea
Taking a derivative of Eq.~\eref{eq:PQz2} and using Eq.~\eref{eq:dPz}, we get,
\bea
\fl \qquad z(1-z) \tilde Q''(z) + \left[\frac{1- \beta}{2} - \frac 12(5 - 3\beta)z \right] \tilde Q'(z) -\left(1- \frac \beta2 \right)\left(\frac 12 - \beta\right) \tilde Q(z) =0\label{eq:Qz}
\eea
Clearly, this is also a hypergeometric differential equation of the form \eref{eq:Hyp}, but with a different parameter set,
\bea
a_2 = 1- \frac \beta 2 = a_1, \quad b_2 = \frac 12 -\beta = b_1 -1, \quad c_2 = \frac{1-\beta}2 = c_1 -1.
\eea

\subsection{Position distribution for $\beta \ne 1$}

\begin{figure}[t]
\centering \includegraphics[width=10.3 cm]{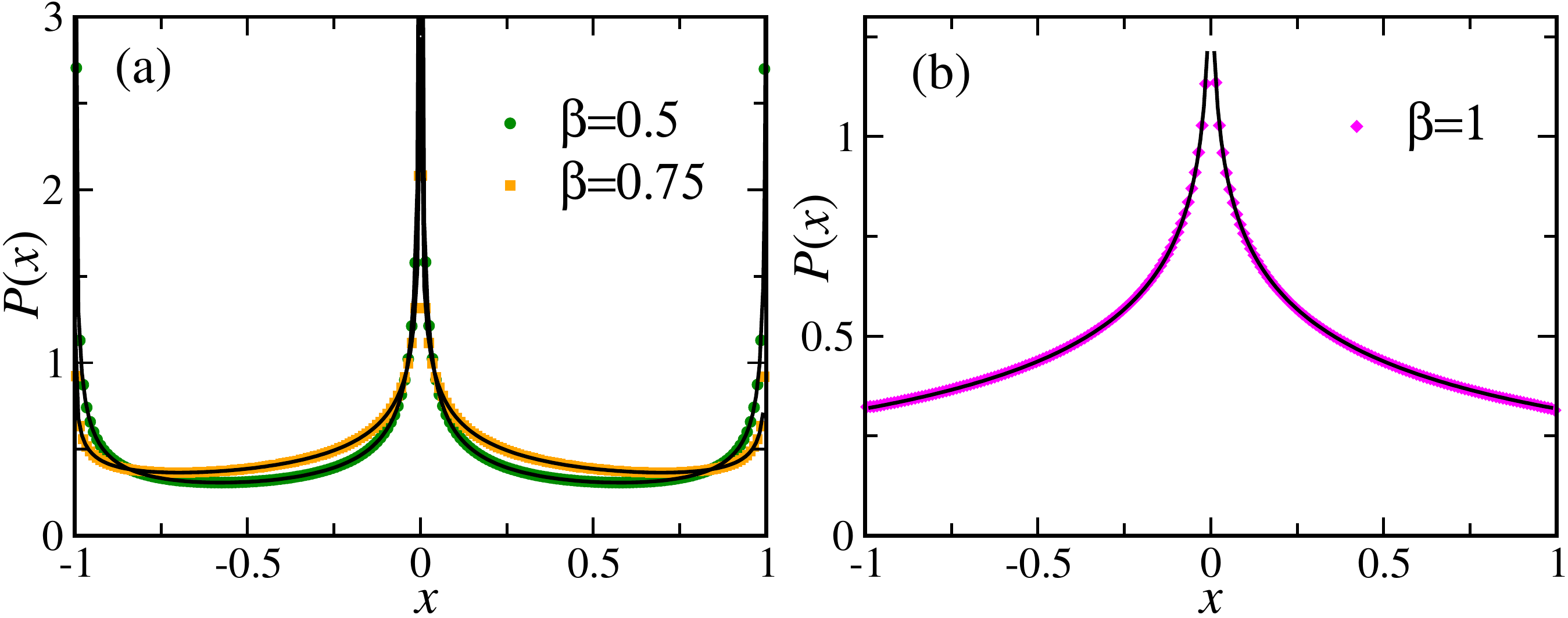} \includegraphics[width=5.1 cm]{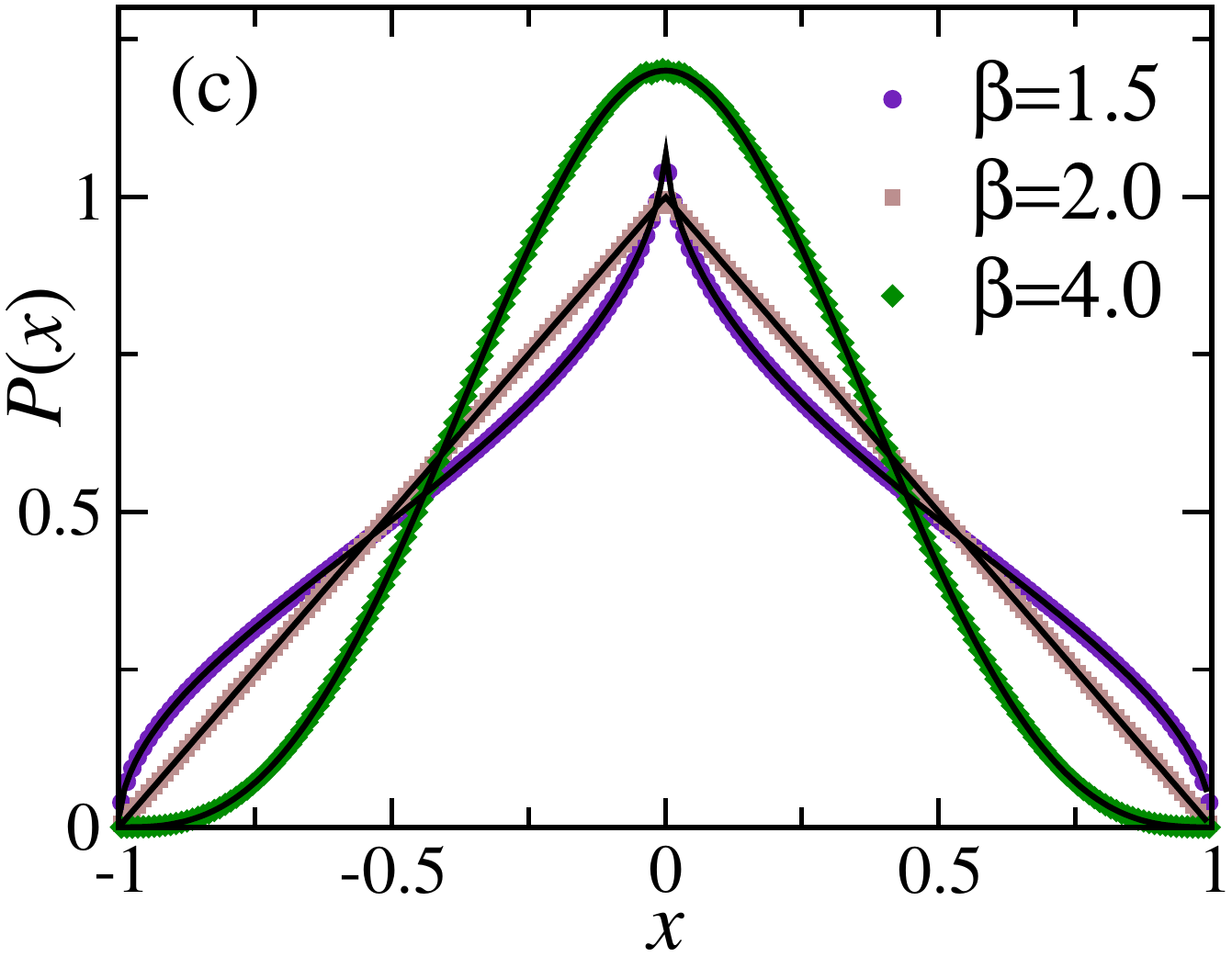}
 \caption{Stationary position distribution $P(x)$ as a function of $x$ for the  3-state model for (a) $\beta <1$, (b) $\beta=1,$  and (c) $\beta>1.$  Here $v_0=1$ and $\mu=1.$ The symbols correspond to the data obtained from numerical simulations while solid lines are obtained from the exact result [see Eq.~\eref{eq:Px_sol} and Eq.~\eref{eq:Px_a1}].}\label{fig:px}
 \end{figure}

The general solutions for Eqs.~\eref{eq:Pz} and \eref{eq:Qz} can be written in terms of the hypergeometric function  $_2 F_1(a,b,c;z)$ \cite{dlmf}.  For $c_1 \ne 1,$ \ie, for $\beta \ne 1,$ these general solutions read,
\bea
\fl \quad P(z) &=&  A_1~\left[ _2F_1(a_1,b_1,c_1;z)\right] + B_1 z^{1-c_1}~ \left[_2F_1(a_1-c_1+1,b_1-c_1+1,2-c_1;z)\right] \label{eq:Pz_sol} \\
\fl \quad Q(z) &=& A_2~ \left[_2F_1(a_2,b_2,c_2;z)\right] + B_2 z^{1-c_2}~ \left[_2F_1(a_2-c_2+1,b_2-c_2+1,2-c_2;z)\right ] \label{eq:Qz_sol} 
\eea
where $A_1, A_2,B_1,B_2$ are arbitrary constants. The case $\beta=1$ is special, which we discuss later. To determine the  constants $A_1, A_2,B_1,B_2$, we first use the original first order equations \eref{eq:PQz1} and \eref{eq:PQz2} which must be satisfied by the solution. Substituting Eqs.~\eref{eq:Pz_sol} and \eref{eq:Qz_sol} in Eq.~\eref{eq:PQz2} and using well known identities involving the hypergeometric function, we get, $B_2 = \frac{B_1}{1 + \beta}$ and $A_2 = \frac{A_1(1-\beta)}{1- 2 \beta}.$
Next, we impose the boundary condition \eref{eq:bc_z}. Once again, using properties of hypergeometric functions, we get
\bea
B_1 = \frac{2 A_1}{\sqrt{\pi}} \frac{\Gamma(\frac{3-\beta}2)\Gamma(\frac 12 +\beta)}{(1- 2 \beta)\Gamma(\frac {1+\beta}2)} \label{eq:B1}
\eea
To completely specify $\tilde P(z)$ we still need $A_1$ which can be determined using the normalization condition,
\bea
\int_{x_-}^{x_+}\id x ~ P (x) =1 \Rightarrow \int_{0}^{v_0/\mu}\id x ~ \tilde P \left[\left(\frac{\mu x}{v_0}\right)^2 \right] =\frac 12.  \label{eq:norm}
\eea
Fortunately, this integral can be performed analytically and yields,
\bea
\fl A_1 = \frac{\mu}{2 v_0}\left[~_3F_2\left({\frac 12 \;\; \frac 32 - \beta \;\; 1-\frac \beta 2 \atop \frac 32 ~~ \frac{3-\beta}2}; 1  \right) - \frac{1}{\beta \sqrt{\pi}} \frac{\Gamma(\frac{3-\beta}{2})\Gamma(\beta -\frac 12)}{\Gamma(\frac{1+\beta}2)}~ _3F_2\left({\frac 12 \;\; 1 - \frac \beta 2 \;\; \frac \beta 2 \atop \frac {1+\beta}2 ~~ \frac{\beta}2+1}; 1  \right) \right]^{-1}\;\; \label{eq:A1}
\eea
where $_pF_q({a_1, a_2, \dots a_p \atop b_1, b_2, \dots b_q};z )$ denotes the generalized hypergeometric function \cite{dlmf}. Finally, we can write an explicit expression for the stationary position probability distribution,
\bea
\fl P(x) &=& A_1~\left [ _2F_1\left(1- \frac \beta 2,\frac 32 - \beta,\frac{3-\beta}2;\left(\frac{\mu x}{v_0}\right)^2 \right) \right. \cr
\fl &&\left. +\frac{2}{\sqrt{\pi}} \frac{\Gamma(\frac{3-\beta}2)\Gamma(\beta+\frac 12)}{(1- 2 \beta)\Gamma(\frac {\beta+1} 2)}\left(\frac{\mu x}{v_0}\right) ^{\beta -1}~ _2F_1 \left(\frac 12,1-\frac \beta 2,\frac{\beta +1}2;\left(\frac{\mu x}{v_0}\right)^2\right) \right] \label{eq:Px_sol}
\eea
where the normalization constant $A_1$ is given by Eq.~\eref{eq:A1}. Note that, $P(x)$ is an even function of $x$ and it depends on the flip rate $\gamma$ comes through the ratio $\beta=\gamma/\mu$ only.  $P(x)$ takes particularly simple form for certain specific values of $\beta,$
\bea
P(x) = \left \{\begin{array}{ccc}
              \frac{\Gamma(\frac 34)}{\sqrt{\pi} \Gamma(\frac 14)} \frac{\sqrt{\mu v_0}}{\sqrt{|x|(v_0^2-\mu^2x^2)}} \qquad \qquad\qquad \qquad & \textrm{for}&  \beta= \frac 12  \cr
                            \frac{\mu}{v_0}(1-\frac{\mu |x|}{v_0}) \qquad \qquad \qquad \qquad \quad \qquad & \textrm{for}& \beta= 2 \cr
              \frac {6\mu}{5v_0} \left[1- 5 (\frac{\mu x}{v_0})^2 - \left(\frac{\mu |x|}{v_0}\right)^3\left((\frac{\mu x}{v_0})^2-5 \right) \right] & \textrm{for}& \beta= 4.
               \end{array}
            \right.\label{eq:Px_simple}
\eea
One can also write an explicit expression for $Q(x)$ using Eqs.\eref{eq:Qz_sol} and \eref{eq:B1},
\bea
\fl Q(x) &=& \frac{A_1}{(1-2\beta)}\left [(1-\beta)~_2F_1\left(1- \frac \beta 2,\frac 12 - \beta,\frac{1-\beta}2;\left(\frac{\mu x}{v_0}\right)^2 \right) \right. \cr
\fl &&\left. + \frac{2}{\sqrt{\pi}} \frac{\Gamma(\frac{3-\beta}2)\Gamma(\beta+ \frac 12)}{(1+\beta)\Gamma(\frac {\beta+1} 2)}\left(\frac{\mu x}{v_0}\right)^{\beta +1}~ _2F_1 \left(\frac 12,1-\frac \beta 2,\frac{\beta +3}2;\left(\frac{\mu x}{v_0}\right)^2\right) \right]. \label{eq:Qx_sol}
\eea
From Eqs.~\eref{eq:Px_sol} and \eref{eq:Qx_sol} and using the relation \eref{eq:PR} between $P(x)$ and $R(x)$ we can also calculate $P_i(x)$ individually in a straightforward manner. However, we do not give explicit expressions for them here. 
Figure \ref{fig:px}(a) and (c) show plots of $P(x)$ as a function of $x$ for different values of $\beta$ calculated from Eq.~\eref{eq:Px_sol} along with the data obtained from numerical simulations. It appears that, similar to the 2-state RTP, the distribution shows two different behaviours near the boundary $x=x_\pm$ depending on the value of $\beta.$ Moreover, it appears from the plots that for $\beta<1,$ $P(x)$ also diverges near the origin $x=0$ while it shows a cusp-like behaviour for large $\beta.$ In the following we investigate the behaviour of $P(x)$ in more details and characterise this change in shape. \\

\noindent{\bf Behaviour near $x=0$:} To understand the behaviour of $P(x)$ near the origin we use the series expansion of the hypergeometric function $_2F_1(a,b,c;z)$ near $z=0,$
\bea
_2F_1(a,b,c;z) = 1+ \frac{ab}{c}z+\frac{ab(1+a)(1+b)}{2c(1+c)} z^2+ \cal O(z^3)
\eea
Using this expansion in Eq.~\eref{eq:Px_sol}, we have, near $x=0,$
\bea \label{asympt_x0}
P(x) \sim \left \{ \begin{array}{l}
                   B_1 \left( \frac{\mu}{v_0} x\right)^{\beta-1} \quad \qquad \quad \textrm{for} \;\;  \beta < 1 \cr
                   A_1(1 - C_1 x^{\beta -1}) \quad \textrm{for} \;\; 1 < \beta < 3 \cr
                   A_1(1 - C_2 x^2) \qquad \textrm{for} \;\; \beta > 3
                   \end{array}
\right.
\eea
where $B_1$ and $A_1$ are given respectively in Eqs. \eref{eq:B1} and~\eref{eq:A1} while $C_1$ and $C_2$ are given~by
\begin{eqnarray}
&&C_1 = \frac{2}{\sqrt{\pi}} \frac{\Gamma(\frac{3-\beta}2)\Gamma(\beta+\frac 12)}{(2 \beta-1)\Gamma(\frac {\beta+1} 2)} \left( \frac{\mu}{v_0}\right)^{\beta-1}\label{C1}  \;, \\
&&C_2  = \frac{2 \beta^2 + 6 - 7\beta}{2(\beta-3)} \left( \frac{\mu}{v_0}\right)^{2} \label{C2} \;.
\end{eqnarray}
Clearly, for $\beta <1,$ $P(x)$ diverges near the origin whereas for $\beta > 1$ it approaches a finite value. The approach also depends on the value of $\beta:$ for $1 < \beta < 3,$ $P(x)$ has a cusp-like behaviour near the origin while for $\beta \ge 3$ it shows a quadratic behaviour, resembling a Gaussian around the origin. Indeed, in the diffusive limit, when $\gamma \to \infty$ and $v_0 \to \infty$ keeping $v_0^2/(2 \gamma) = D$ fixed (as a consequence $\beta \to \infty$ in this limit), we find from Eq. \eref{C2} that $C_2 \to \mu/(2D)$. As a result, from the third line of \eref{asympt_x0}, we recover the Boltzmann distribution $P(x) \sim e^{-\mu/(2D) x^2}$ which actually holds for all $x$. \\

\noindent{\bf Behaviour near $x=x_\pm$:}  The position distribution $P(x)$  also shows an interesting behaviour near the boundaries $x= x_{\pm}.$ As $P(x)$ is symmetric in $x,$ it suffices to explore its nature near one boundary, say $x_+.$ To characterise the same we use the series expansion of $_2F_1(a,b,c;z)$ near $z=1.$ From Eq.~\eref{eq:Pz_sol}, we have, for $z\to 1^-,$
\bea
\tilde P(z) \sim \left\{ \begin{array}{l}
                   (1-z)^{\beta -1} \quad \textrm{for} \;\;  \beta < 3 \cr
                    (1-z)^2  \quad \textrm{for} \;\; \beta > 3 \cr \;.
                   \end{array}
                   \right.
                   \eea
Hence, near $x=x_+,$ we have the following behaviour of $P(x):$
\bea
P(x) \sim \left\{ \begin{array}{l}
(x_+ - x)^{\beta -1} \quad \textrm{diverges for} \;\;  \beta < 1 \cr
                    (x_+ - x)^{\beta -1}  \quad \textrm{vanishes for} \;\; 1< \beta \le 3 \cr
                    (x_+ - x)^2  \quad \textrm{vanishes for} \;\; \beta > 3 \;.
                   \end{array}
                   \right.
                  \eea
A similar behaviour is seen also near $x=x_-$. Note that this ``freezing'' for the leading behaviour for $\beta > 3$ occurs only for the three-state
model, but not for the two-state model \cite{RTP_trap}.

%We discuss this scenario separately in the following. 

\subsection{Position distribution for  $\beta=1$} 

As mentioned before, the case $\beta=1$ is special. In this case,
the differential equations~\eref{eq:Pz} and \eref{eq:Qz} reduce to,
\bea
z(1-z) \tilde P''(z) + (1-2z) \tilde P'(z) - \frac 14 \tilde P(z) =0 \label{eq:Pz_al1}\\
z(1-z) \tilde Q''(z) -z \tilde Q'(z) + \frac 14 \tilde Q(z) =0. \label{eq:Qz_al1}
\eea
which correspond to two hypergeometric equations with $c_1=1$ and $c_2=0,$ along with  $a_1=a_2=b_1=1/2,b_2=-1/2.$ Eq.~\eref{eq:Pz_sol} is not a general solution anymore as the two hypergeometric functions therein become identical.  We use Mathematica to solve Eqs.~\eref{eq:Pz_al1} and \eref{eq:Qz_al1} and it turns out that the general solutions can be expressed in the form,
\bea
\tilde P(z) &=&  \frac {2 A_1} \pi K(1-z) + B_1 \cal Q_{-\frac 12}(2z-1) \label{eq:Pz_al1_sol}\\
\tilde Q(z) &=& A_2 z ~_2F_1\left(\frac 12, \frac 32, 2;z \right) + B_2 ~G^{20}_{22}\left({\frac 12~ \frac 32 \atop 0 ~1};z \right). \label{eq:Qz_al1_sol}
\eea
Here $K(u)$ is the Legendre's complete elliptic integral of the first kind (see Ref.~\cite{Gradshteyn} and Eq.~19.2.8 in Ref.~\cite{dlmf}), $G^{mn}_{pq}({a_1,\dots a_p \atop b_1 \dots b_q};z)$ is the Meijer's G-function (see Ref.~\cite{Gradshteyn} and Eq.~16.17.1 in Ref.~\cite{dlmf}) and $\cal Q_\nu(u)$ is the Legendre function of the second kind (see Eq.~14.3.7 in Ref.~\cite{dlmf}). 

To determine the arbitrary constants $A_1,A_2,B_1$ and $B_2$ we use the same strategy as in the previous section. First, we note that the solutions in Eqs.~\eref{eq:Pz_al1_sol} and \eref{eq:Qz_al1_sol} must satisfy the original first order equations \eref{eq:PQz1} and \eref{eq:PQz2} with $\beta=1$  for all values of $z.$ We then look at the behaviour of $\tilde P(z)$ and $\tilde Q(z)$ in Eqs.~\eref{eq:Pz_al1_sol} and \eref{eq:Qz_al1_sol} near $z=0$. In this limit both $K(1-z)$ and  $G^{20}_{22}\left({\frac 12~ \frac 32 \atop 0 ~1};z \right)$ diverge logarithmically whereas the Legendre and hypergeometric functions approach a constant value. 
Substituting the series expansions of these functions back  into Eq.~\eref{eq:PQz2} and comparing coefficients of $\ln z$ and different powers of $z,$ we get, $B_2 = A_1,$ and $\quad A_2 = -\frac \pi 4 B_1.$ It is also straightforward to check that Eq.~\eref{eq:PQz1} gives the same relation. We still have two independent constants $A_1$ and $B_1.$  To determine these we use the boundary condition \eref{eq:bc_z}.  Using the limiting behaviours of the special functions we have, for $z \to 1^{-},$ $\tilde P(z) - \tilde Q(z) = B_1 + \cal O(1-z)$ which immediately implies $B_1=0$ [see Eq.~\eref{eq:bc_z}].
% Using the limiting behaviour,
% \bea
% \lim_{z \to 1^-} G^{20}_{22}\left({\frac 12~ \frac 32 \atop 0 ~1};z \right) =1
% \eea 
The last remaining constant $A_1$ can be determined from the normalization condition \eref{eq:norm} and yields $A_1=\frac{\mu}{\pi v_0}.$
Finally, we have, for $\beta=1,$
\bea
P(x) = \frac {2 \mu}{\pi^2 v_0} K\left (1- \frac{\mu^2 x^2}{v_0^2} \right),\quad \textrm{and} \quad  Q(x) = \frac{\mu}{\pi v_0} G^{20}_{22}\left({\frac 12~ \frac 32 \atop 0 ~1}; \frac{\mu^2 x^2}{v_0^2}  \right). \label{eq:Px_a1}
\eea
Figure \ref{fig:px}(b) shows a plot of $P(x)$ for $\beta=1$ together with the same obtained from numerical simulations. 
To  understand the behaviour near the origin $x=0$ and the boundaries $x=x_\pm$ we look at the series expansion of $P(x).$ Near $x=0,$ a logarithmic divergence is seen, $P(x) \sim - \ln x.$
On the other hand, near the boundaries $x=x_\pm,$ $P(x)$ approaches a constant value, $\lim_{x \to x_\pm} P(x) = \frac{\mu}{\pi v_0}.$

% 
% 
% The general solution of Eq.~\eref{eq:Pz} then takes a different form \cite{Gradshteyn},
% \bea
% \fl \qquad \tilde P(z) &=& A_1 ~_2F_1(\frac 12,\frac 12,1;z)+ B_1 \bigg[~_2F_1(\frac 12,\frac 12,1;z) \ln z + \sum_{k=1}^\infty \omega_k z^k \bigg] \cr
% \fl \textrm{with} \;\; \omega_k &=& 2\left(\frac{(\frac 12)_k}{k!}\right)^2 \left[\psi \left(k+\frac 12 \right)-\psi\left(\frac 12\right) - \psi(k+1)+\psi(1) \right].
% \eea
% Here $(a)_k$ denotes the Pochhammmer symbol and $\psi(u)= \frac{\id}{\id u} \ln \Gamma(u)$ is the digamma function. 

\section{Conclusion}

In this paper, we have solved exactly the stationary position distribution of a one-dimensional run-and-tumble (RTP) particle with three discrete internal states and subjected to an external harmonic potential. To our knowledge, this is the first exact solution with three states that generalizes the well-known result for the standard two-state RTP. We showed that the stationary state exhibits a rich behavior as a function of the single parameter $\beta = \gamma/\mu$ (where $\gamma$ represents the rate at which the internal state changes and $\mu$ is the stiffness of the trap). One of the interesting outcomes is that the stationary distribution undergoes a shape-transition at $\beta=1$.

While we were able to characterise the stationary state of a three-state RTP in a harmonic trap exactly, it would be interesting
to study the relaxational dynamics towards this stationary state, as was recently done for the two-state RTP \cite{RTP_trap}.  
It would also be natural to extend our studies to non-harmonic potentials, such as $U(x) \sim |x|^p$, with $p>0$. Another natural extension 
would be to consider an RTP particle with more than $3$ internal states. Finding even the stationary state of a general $n$-state RTP with $n > 3$ remains a challenging open problem.

\ack
We acknowledge support from the project 5604-2 of the Indo-French Centre for the Promotion of Advanced Research (IFCPAR). 
S. N. M. acknowledges the support from the Science  and  Engineering  Research  Board  (SERB,  government  of  India)  under  the  VAJRA  faculty  scheme  (Ref.VJR/2017/000110) during a visit to the Raman Research Institute in 2019, where part of this work was carried out. U. B. acknowledges support from Science and Engineering Research Board (SERB), India under Ramanujan Fellow-ship (Grant No.  SB/S2/RJN-077/2018) and CNRS for a one month visit to LPTMS, Univ. Paris-Sud.

\end{document}